 \documentclass[
twocolumn,showpacs, amsmath, amssymb, floatfix,
nofootinbib,wrapfloat byrevtex]{revtex4}

 \usepackage{amsmath,mathrsfs,graphicx,epstopdf,placeins,float}
 \usepackage{booktabs}
 \usepackage{multirow}
\usepackage{color}

 \newcommand{\beq}[1]{\begin{equation}\label{#1}}
 \newcommand{\eeq}{\end{equation}}
 \newcommand{\bea}[1]{\begin{eqnarray}\label{#1}}
 \newcommand{\eea}{\end{eqnarray}}
 \newcommand\figcaption{\def\@captype{figure}\caption}
 \newcommand\tabcaption{\def\@captype{table}\caption}

\begin{document}

 \title{Geometric description of the Schr\"{o}dinger equation \\
 in $3n+1$ dimensional configuration space}
 \author{M. Abdul Wasay $^{\!\!1}$}
\email{muhammad.wasay@uaf.edu.pk}

\author{Asma Bashir$^{1,}$}
\author{Benjamin Koch$^{2,}$}
\email{bkoch@fis.puc.cl}
\author{Abdul Ghaffar$^{1,}$}
\affiliation{$^1$Department of Physics, University of Agriculture, Faisalabad 38040, Pakistan;\\
$^2$Instituto de F\'{i}sica, Pontificia Universidad Cat\'{o}lica de Chile,\\
Av. Vicu\~{n}a Mackenna 4860,\\
782-0436 Macul, Santiago, Chile}
 \begin{abstract}
 We show that for non-relativistic free particles, the (bosonic) many particle equations
 can be rewritten in geometric fashion in terms of a classical theory of conformally stretched spacetime.
 We further generalize the results for the particles subject to a potential.
  \end{abstract}

 \maketitle
 \smallskip

\section{Introduction}

In general relativity (GR), gravitational interactions and phenomena
are formulated and understood in terms of purely geometrical quantities.
In this theory, mass- and energy-sources geometrically curve spacetime and
point particles move on simple geodesics in this background.
This beautiful and intuitive geometric fundament (apart from its great phenomenological success)
are the reason for GRs unbroken popularity, more than 100 years after its formulation.

Quantum mechanics (QM) on the other hand deals with more abstract objects such as
wave functions or probability amplitudes, which are usually less familiar
than the trajectories and curved surfaces that appear in GR.
It is thus natural that there are numerous attempts
to reformulate quantum mechanics in a more geometrical way \cite{Santamato:1984qe, geometry8,Shojai:2000us,Carroll:2004hs,Carroll:2007zh,Abraham:2008yr,Koch:2008rh,Koch:2010bz,Koch:2010ija,Nicolini:2010nb,Carroll:2010ef,Isidro:2010nc,Koch:2011hc,Acosta:2011rb,Mehdipour:2011mc,Biro:2013aaa,geometry10,geometry11,KOCH:2013eoa,Arkani-Hamed:2013jha}.

In order to cast quantum mechanics in the geometric language of GR
one typically needs to define physical trajectories and a background space.
Using the language of \cite{Bohm,Bohm2}, it was shown that
such trajectories naturally arise in the configuration space
for the complex Klein Gordon equation.
It was further found that the evolution equation for those trajectories
can be cast in the form of a geodesics equation in a conformally rescaled
configuration space \cite{KOCH:2013eoa,Koch:2010ija,geometry8,geometry11,Koch:2008rh,Shojai:2000us}.
Thus, the relativistic Klein Gordon equation can be rewritten
in a geometric language with non-trivial trajectories in configuration space.

The purpose of this article is to discuss the non-relativistic limit of those findings
and to examine to which extent one can accommodate a quantum mechanical potential
in this reformulation.

The paper is organized as follows:
in section II we examine the non-relativistic limit of the quantum Klein-Gordon equation and for a system of free bosonic particles,
which is rewritten by splitting the wave function in amplitude and phase.
In section III, a geometric theory using an $n$-particle action is presented and it is shown that this geometric formulation
can (by the use of suitable matching conditions)
be identified with the rewritten quantum mechanics given in section II.
In section IV, the results are generalized to case where the particles are not free but are subject
 to a potential. Section V contains some useful discussions and section VI summarizes the results.

\section{\textbf{Non-Relativistic Limit of the Klein-Gordon equation}}

In this section it is shown how the non relativistic limit of the quantum Klein Gordon
can be rewritten by splitting the wave function in amplitude and phase.
The notation will largely follow \cite{KOCH:2013eoa}.
In the non-relativistic limit space and time are not on equal footing so here spatial and temporal derivatives are treated differently.
The quantum Klein-Gordon equation. is
\bea{}
\left(\Box+\frac{M^2c^2}{\hbar^2} \right)\psi=0.
\label{first}
\eea
To get non-relativistic limit of \eqref{first} we make use of the following ansatz
\bea{}
\psi(x,t)=\phi(x,t)\textmd{exp}\left(\frac{-iMc^2t}{\hbar}\right).
\label{second}
\eea
In the non-relativistic limit the energy difference between total energy and rest mass is assumed to be small such that
\bea{}
E^\prime=E-Mc^2,\nonumber
\eea
with
\bea{}
E^\prime\ll\ Mc^2.\nonumber
\eea
These approximations yield the Schr\"{o}dinger equation for spinless particles. The
many particle Schr\"{o}dinger equation can be written as
\bea{}
\left(\sum\limits_{j=1}^n\frac{\hbar}{2M_j}\partial_j^m\partial_{jm}+i\partial_0\right)\psi(x_1,x_2,...,x_n,t_1)=0.
\label{third}
\eea
Here, $M$ is the single particle mass and $M_j$ represents $n$-particle mass. The particle to be affected out of $n$-particles is denoted by the index $j$ and $m$ is the space index in three flat dimensions.
One factorize the wave function into amplitude and phase as just like it is done in \cite{Bohm,Bohm2}
\bea{}
\psi=Pe^{iS/\hbar}\nonumber.
\eea

Since we are working in non-relativistic limit, so "time" is absolute, and we have only one time coordinate $t_1$ instead of $n$ time coordinates $t_j$. The quantum phase then reads
\bea{}
S(t_j, \vec{x}_j)=-Mt_1+\tilde S(t_1,\vec{x}_j)\nonumber.
\eea
The same projection onto a single time coordinate
is true for Hamilton's principle function $S_H$, and the amplitude~$P$
\bea{}
P(t_j,\vec{x}_j)=P(t_1,\vec{x}_j)\nonumber.
\eea
Using the above definition of amplitude and phase in non-relativistic limit we rewrite the wave function as
\bea{}
\psi(t_1,\vec{x}_j)=P(t_1,\vec{x}_j)e^{i\left(\tilde S(t_1,\vec{x}_j)-Mt_1\right)/\hbar}.
\label{four}
\eea
Thus, (\ref{third}) reads
\bea{}
\left(\sum\limits_{j=1}^n\frac{\hbar}{2M_j}\partial_j^m\partial_{jm}+i\partial_0\right)P(t_1,\vec{x}_j)e^{i\left(\tilde S(t_1,\vec{x}_j)-Mt_1\right)/\hbar}=0.\nonumber
\eea

Here, $\partial P/\partial t_1=0$, as for large $t$ the amplitude on the average is zero.
Taking the real part of the above equation after using Taylor series, one gets
\bea{}
\sum\limits_{j=1}^n\frac{\hbar^2}{2M_j}\frac{\partial_j^m\partial_{jm}P(t_1,\vec{x}_j)}{P(t_1,\vec{x}_j)}=\sum\limits_{j=1}^n\frac{(\partial_{jm}\tilde S)(\partial_j^m\tilde S)}{2M_j}+\dot{\tilde S}-M,\nonumber
\eea
where
the derivative with respect to $t_1$ is represented by a dot.
Thus, one can write
\bea{}
Q=\sum\limits_{j=1}^n\frac{(\partial_{jm}\tilde S)(\partial_j^m\tilde S)}{2M_j}+\dot{\tilde S}-M,
\label{five}
\eea
where,
\bea{}
Q=\sum\limits_{j=1}^n\frac{\hbar^2}{2M_j}\frac{\partial_j^m\partial_{jm}P(t_1,\vec{x}_j)}{P(t_1,\vec{x}_j)}.\nonumber
\eea

Equation \eqref{five} can be thought of as Hamilton-Jacobi equation as in classical mechanics.
The term $Q$ here represents the quantum potential $Q(x_1,x_2,...,x_n,t_1)$ and $P(x_1,x_2,...,x_n,t_1)$ is the pilot wave \cite{Bohm,Bohm2}.
The defined non relativistic wave function allows to construct conserved current as
\bea{}
\partial_0(\psi^*\psi)-\sum\limits_{j=1}^n\partial_j^m\left(\frac{i\hbar}{2M_j}(\psi^\ast\overleftrightarrow{\partial}_{jm}\psi)\right)=0\nonumber.
\eea

The conserved current with the defined non-relativistic wave function yields a real equation
\bea{}
\partial_0(P^2)+\sum\limits_{j=1}^n\partial_{jm}\left(\frac{P^2}{M_j}(\partial_j^m\tilde S)\right)=0.
\label{six}
\eea
Equation \eqref{six} is the second equation representing the conserved current.
Notice that $\partial_0(P^2)$ is not zero because here negative changes in amplitude would also become positive.

The interpretation in terms of trajectories is introduced by defining velocities and momentum by
\bea{}
p_j^m=M_j\frac{dx_j^m}{ds}=\partial_j^m\tilde S.
\label{seven}
\eea

Now the equation of motion for $n$ non-relativistic particles is given by
\bea{}
\frac{dx_j^m}{ds}=\frac{\partial_j^m\tilde S}{M_j}\nonumber.
\eea
Using
\bea{}
\frac{d}{ds}=\sum\limits_{j=1}^n\frac{d}{dx_j^m}\frac{dx_j^m}{ds},\nonumber
\eea
one finds
\bea{}
\frac{d^2x_j^m}{ds^2}=\sum\limits_{i=1}^n\frac{(\partial_i^l\tilde S)(\partial_j^m\partial_{il}\tilde S)}{M_j^2}.
\label{eight}
\eea
Here, $ds=dt_1$ is the common time coordinate for all $n$ particles.
Please note that those equations are just a rewriting of the initial equation (\ref{third}).
Please note further that in addition to analogous manipulations,
the de Broglie Bohm (dBB) interpretation consists in associating physical reality to the trajectories (\ref{eight}).
Since the dBB theory is deterministic and non-local, equation \eqref{eight} confirms the non-local nature of dBB theory i.e., position of one particle depends on the position of all other particles constituting the system. This can be thought of as Newtons's second law of motion and this equation shows that the motion of jth particle is affected by a 'force' $\sum\limits_{i=1}^n\frac{(\partial_i^l\tilde S)(\partial_j^m\partial_{il}\tilde S)}{M_j^2}$ which involves the position of all the particles \cite{geometry4}.\\
However, the purpose of this article is not to assign physical reality to the laws of quantum motion.
Instead the purpose is to show that quantum mechanics can be rewritten in a geometric language, independent of the question
of physical reality of the appearing trajectories.
In the equations \eqref{five}-\eqref{eight},
the amplitude $P$, and the phase $S$, and  $Q$ depend on $1+3n$ coordinates,
$3n$ of space and $1$ of time.
In order to abbreviate notation one can define
\bea{}
x^L=(t_1,\vec{x_1},\vec{x_2},...,\vec{x_n}),\nonumber
\eea
such that $\partial_j^m\rightarrow\partial^L$ and $\partial_{jm}\rightarrow\partial_L$; with
\bea{}
Q=\frac{\hbar^2}{2M_j}\frac{\partial^L\partial_LP(t_1,\vec{x_j})}{P(t_1,\vec{x_j})}.\nonumber
\eea
As mentioned above,we will take the temporal derivative of amplitude as zero as on the average $\partial P/\partial t_1=0$. So there remain only spatial derivatives.

Thus the non-relativistic limit of (\ref{third}) is equivalently given by the following set of equations
\bea{}
Q=\frac{(\partial^L\tilde S)(\partial_L\tilde S)}{2M_j}+\dot{\tilde S}-M,
\label{nine}
\eea
\bea{}
\partial_L\left(\frac{P^2(\partial^L\tilde S)}{M_j}\right)+\partial_0(P^2)=0,
\label{ten}
\eea
\bea{}
p^L=M_j\frac{dx^L}{ds}=\partial^L\tilde S,
\label{eleven}
\eea
\bea{}
\frac{d^2x^L}{ds^2}=\frac{(\partial^N\tilde S)(\partial^L\partial_N\tilde S)}{M_j^2}.
\label{twelve}
\eea
It is the purpose of the following section to show that
this set of equations can be obtained from a purely geometrical
formulation in configuration space.

\section{\textbf{Geometry of Configuration Space}}

We consider a $1+3n$ dimensional configuration space of $n$-particles with one single time coordinate. The coordinates are denoted by
\bea{}
\hat{x}^\Lambda=(\hat{t}_1,\hat{\vec{x}}_1,\hat{\vec{x}}_2,...,\hat{\vec{x}}_n)\nonumber.
\eea

The model specifying the curvature of such space uses a single $1+3n$ dimensional scalar equation given by
\bea{}
P_s\left(\hat{R}+k\hat{L}_M\right)=\hat{R}+k\hat{L}_M.
\label{thirteen}
\eea
Here $P_s$ is symmetrization operator between different particles $x_i^\lambda$ and $x_j^\lambda$, $\hat{R}$ is Ricci scalar, $\hat{L}_M$ is matter lagrangian and $k$ is coupling constant representing the extent of interaction between particles and field.

With the symmetrization condition \eqref{thirteen} the particle action reads
\bea{}
S\left(\hat{g}_{\Lambda\Delta}\right)=\int dt\int dx^{3n}\sqrt{\left|\hat{g}\right|}\left(\hat{R}+k\hat{L}_M\right)
\label{fourteen}
\eea

The local conformal part of the theory is described separately by splitting the metric $\hat{g}$ into a conformal function $\phi(\vec{x}_j,t_1)$and a flat part $\eta$ \cite{KOCH:2013eoa}. The conformal transformation here is given by
\bea{}
\hat{g}_{\Lambda\Gamma}=\phi^{\frac{4}{3n-1}}\eta_{LG}.
\label{fifteen}
\eea
The inverse of the metric is given by
\bea{}
\hat{g}^{\Lambda\Gamma}=\phi^{\frac{-4}{3n-1}}\eta^{LG}.
\label{sixteen}
\eea

The lower Greek and lower Roman index are identified as $\hat{\partial}_\Lambda=\partial_L$ so that the adjoint derivatives are different in each notation i.e.,
\bea{}
\hat{\partial}^\Lambda&=&g^{\Lambda\Sigma}\hat{\partial}_\Sigma=\phi^{\frac{-4}{3n-1}}\eta^{LS}\partial_S\nonumber
\\
\hat{\partial}^\Lambda&=&\phi^{\frac{-4}{3n-1}}\partial^L\nonumber
\\
\hat{\partial}_\Lambda&=&\phi^{\frac{4}{3n-1}}\partial_L\nonumber.
\eea

\subsection{Geometric Dual of the 1st Equation:}

The particle action in terms of $\phi$ and $g_{LD}$ is \cite{KOCH:2013eoa}
\bea{}
S\left(\phi,g_{LD}\right)=\int dt\int dx^{3n}\sqrt{|g|}~[\frac{12n}{1-3n}(\partial^L\phi)(\partial_L\phi)\nonumber.
\\
+\phi^2(R+kL_M)]\nonumber
\eea
It is interesting to note that in a truly gravitational theory a similar factorization
of the conformal factor can be used to mimic the effects of dark matter \cite{geometry14}.
However, in this toy model in configuration space
we are only interested in studying flat ``Minkowski'' background space so $g_{LG}=\eta_{LG}$ and $|g|=1$ and $R=0$. So the action simplifies to
\bea{}
S\left(\phi\right)=\int dt\int dx^{3n}\left(\frac{12n}{1-3n}(\partial^L\phi)(\partial_L\phi)+\phi^2(kL_M)\right)\nonumber.
\eea

The equation of motion for $\phi$ is
\bea{}
\frac{12n}{1-3n}\frac{\partial^L\partial_L\phi}{\phi}=kL_M.
\label{seventeen}
\eea

The matter Lagrangian $L_M$ is given by
\bea{}
L_M=\frac{2(\partial^L\tilde{S}_H)(\partial_L\tilde{S}_H)}{2\hat{M}_G}+\frac{\partial\tilde{S}_H}{\partial t_1}-\hat{M}\nonumber,
\eea
which  also contains square of first order temporal derivatives.
The equation of motion for $\phi$ reads
\bea{}
\frac{12n}{k(1-3n)}\frac{\partial^L\partial_L\phi}{\phi}=\frac{2(\partial^L\tilde{S}_H)(\partial_L\tilde{S}_H)}{2\hat{M}_G}+\dot{\tilde{S}}_H -\hat{M}.
\label{eighteen}
\eea

With the following matching conditions
\bea{}
k=\frac{12n}{1-3n}.\frac{2\hat{M}_G}{\hbar^2}\nonumber
\eea
\bea{}
\phi(\vec{x}_j,t_1)=P(\vec{x}_j,t_1)\nonumber
\eea
\bea{}
\tilde{S}_H(\vec{x}_j,t_1)=\tilde{S}(\vec{x}_j,t_1)\nonumber
\eea
\bea{}
M_j=\hat{M}_G\nonumber
\eea
we find that  the geometrical equation \eqref{eighteen} is identical to
the quantum equation \eqref{nine}.
Please note that
in relativistic case the dual theory was developed in $4n$ dimensions \cite{KOCH:2013eoa} while here
equation \eqref{eighteen} is obtained from the geometric theory developed in $(1+3n)$ dimensions.
This clearly reflects the fact that here space and time are treated differently.
In contrast to the relativistic case, the non-relativistic case in addition to spatial changes also incorporates temporal changes of the pilot wave.

\subsection{Geometric Dual of the 2nd Equation}

The stress energy tensor is of the matter part given by
\bea{}
T^{\Lambda\Delta}=\frac{2(\hat{\partial}^\Lambda\tilde{S}_H)(\hat{\partial}^\Delta\tilde{S}_H)}{\hat{M}_G}+~~~~~~~~~~~~~~~~~~~~~~~~\nonumber
\\
g^{\Lambda\Delta}\left(\frac{(\hat{\partial}^\Lambda\tilde{S}_H)(\hat{\partial}_\Lambda\tilde{S}_H)}{\hat{M}_G}+
\partial_0\tilde{S}_H-\hat{M}\right)\nonumber,
\eea
where $\partial_0=\partial/\partial t_1$.

Since the stress energy tensor is covariantly conserved, so
\bea{}
\nabla_\Lambda T^{\Lambda\Delta}=0\nonumber
\eea

From this, we can write
\bea{}
\frac{(\hat{\partial}^\Delta\tilde{S}_H)\nabla_\Lambda(\hat{\partial}^\Lambda\tilde{S}_H)}{\hat{M}_G}=0,
\label{ninteen}
\eea
\bea{}
\frac{(\hat{\partial}^\Lambda\tilde{S}_H)\nabla^\Delta(\hat{\partial}_\Lambda\tilde{S}_H)}{\hat{M}_G}=0,
\label{twenty}
\eea
\bea{}
\frac{(\hat{\partial}_\Lambda\tilde{S}_H)\nabla_\Lambda(\hat{\partial}^\Lambda\tilde{S}_H)}{\hat{M}_G}+\nabla_\Lambda(\partial_0\tilde{S}_H)=0.
\label{twenty one}
\eea

The Levi-Civita connection is given by
\bea{}
\Gamma^\Sigma_{\Lambda\Delta}=\frac{1}{2}g^{\Sigma\Xi}\left(\partial_\Lambda g_{\Delta\Xi}+\partial_\Delta g_{\Xi\Lambda}-\partial_\Xi g_{\Lambda\Delta}\right)\nonumber,
\eea
\bea{}
\Gamma^\Sigma_{\Lambda\Delta}=\frac{1}{2}\phi^{\frac{-4}{3n-1}}[(\partial_L\phi^{\frac{4}{3n-1}})\delta^S_D+(\partial_D\phi^{\frac{4}{3n-1}})\delta^S_L\nonumber
\\
-(\partial^S\phi^{\frac{4}{3n-1}})\eta_{LD}]\nonumber.
\eea

With this, the relation \eqref{twenty one} reads
For the first term
\bea{}
\nabla_\Lambda(\hat{\partial}^\Lambda\tilde{S}_H)=\partial_\Lambda(\partial^\Lambda\tilde{S}_H)~~~~~~~~~~~~~~~~~~~~~~~\nonumber
\\
+\frac{1}{2}g^{\Sigma\Xi}[\partial_\Lambda g_{\Delta\Xi}+\partial_\Delta g_{\Xi\Lambda}-\partial_\Xi g_{\Lambda\Delta}]\nonumber
\\
\times [\partial^\Lambda\tilde{S}_H]=0\nonumber,
\eea
or
\bea{}
\phi^{\frac{-2-6n}{3n-1}}\partial_L(\phi^2(\partial^L\tilde{S}_H))=0
\label{twenty two}
\eea
For the second term in \eqref{twenty one} we used
\bea{}
\nabla_\Lambda(\partial_0\tilde{S}_H)=\partial_\Lambda(\partial_0\tilde{S}_H)\nonumber
\eea
and
\bea{}
\nabla_\Lambda(\partial_0\tilde{S}_H)=\partial_0(\partial_L\tilde{S}_H).
\label{twenty three}
\eea
\\
From \eqref{twenty one}, with  \eqref{twenty two} and  \eqref{twenty three}, one gets
\bea{}
\left(\frac{\partial_L\left[\phi^2(\partial^L\tilde{S}_H)\right]}{\hat{M}_G}+\partial_0(\phi^2)\right)=0.
\label{twenty four}
\eea

With the given matching conditions, one confirms that
equation \eqref{twenty four} is identical to \eqref{ten}.

In non-relativistic limit $\psi^*\psi$ can be interpreted as probability density, but it is not possible to provide an interpretation for the probability of Klein-Gordon equation by $\psi^*\psi$.
The probability interpretation of Klein-Gordon equation is given in terms of Klein-Gordon current, that is conserved with respect to time.
Here $\phi(t_1,\vec{x}_j)$ enters two ways, $\phi(t_1,\vec{x}_j)$ represents the interaction of matter with the conformal field and its square
 $\phi^2(t_1,\vec{x}_j)$ represents the probability density.

\subsection{Geometric Dual of the 3rd Equation:}

The momentum is defined by the derivative of the Hamilton principle function
$\tilde{S}_H$ as suggested by Hamilton-Jacobi formalism
\bea{}
\hat{p}^\Lambda=(\hat{\partial}^\Lambda\tilde{S}_H).
\label{twenty five}
\eea
This is identical  to the third equation \eqref{eleven}.

\subsection{Trajectory Equation of Motion}

It remains to obtain the geometric dual of the relation (\ref{twelve}).
One notes that the total derivative is
\bea{}
\frac{d}{d\hat{s}}&=&\frac{d\hat{x}^\Lambda}{d\hat{s}}\hat{\partial}_\Lambda\nonumber
\\
\frac{d}{d\hat{s}}&=&\phi^{\frac{4}{3n-1}}\frac{dx^L}{ds}\partial_L\nonumber
\\
\frac{d}{d\hat{s}}&=&\phi^{\frac{4}{3n-1}}\frac{d}{ds}\nonumber.
\eea

Applying this relation to momenta gives
\bea{}
\hat{M}_G\left(\frac{d\hat{x}^\Lambda}{d\hat{s}}\right)
=(\hat{\partial}^\Lambda\tilde{S}_H)\nonumber,
\eea
and
\bea{}
\frac{d^2\hat{x}^\Lambda}{d\hat{s}^2}=\frac{d}{d\hat{s}}\frac{\left(\hat{\partial}^\Lambda\tilde{S}_H\right)}{\hat{M}_G}\nonumber.
\eea

Using the identity
\bea{}
\frac{d}{d\hat{s}}=\hat{\partial}_\Lambda\frac{{d\hat{x}^\Lambda}}{d\hat{s}}\nonumber,
\eea
one finds
\bea{}
\frac{{d^2\hat{x}^\Lambda}}{d\hat{s^2}}=\frac{\left(\hat{\partial}^\Delta\tilde{S}_H\right)\hat{\partial}_\Delta\left(\hat{\partial}^\Lambda\tilde{S}_H\right)}{\hat{M}_G^2}.
\label{twenty six}
\eea
This is the equation of motion for non-relativistic particles. We can relate it with another equation, the "Geodesic equation of motion" in non-relativistic limit.

\bea{}
\frac{d^2\hat{x}^\Lambda}{d\hat{s}^2}=-\Gamma^\Lambda_{00}
\eea
\bea{}
\frac{d^2\hat{x}^\Lambda}{d\hat{s}^2}=\frac{1}{2}\hat{g}^{\Lambda\Gamma}\frac{\partial \hat{g}_{00}}{\partial \hat{x}^\Lambda}\nonumber.
\eea
Applying a conformal transformation it leads to
\bea{}
M\frac{d^2\hat{x}^\Lambda}{d\hat{s}^2}=-\frac{1}{2}M(\nabla\phi^\frac{4}{3n-1})\nonumber
\eea
which is just like the equation of motion in "Newtonian gravity".
With the given matching conditions, equation \eqref{twenty six} is identical to \eqref{twelve}.

\section{\textbf{Including a Potential}}

The Schr\"{o}dinger equation in presence of potential is
\bea{}
\left(\frac{-\hbar^2}{2M}\nabla^2+V\right)\psi=i\hbar\frac{\partial\psi}{\partial t}\nonumber
\eea
It is straight forward to see that this modifies the previous discussion only
by an additional term in
\bea{}
Q=\sum\limits_{j=1}^n\frac{(\partial_{jm}\tilde S)(\partial_j^m\tilde S)}{2M_j}+\dot{\tilde S}+V-M.
\label{twenty seven}
\eea

\subsection{Geometric Dual}
The particle action in terms of $\phi$ and $g_{LD}$ is
\bea{}
S\left(\phi,g_{LD}\right)=\int dt\int dx^{3n}\sqrt{|g|}~[\frac{12n}{1-3n}(\partial^L\phi)(\partial_L\phi)+\nonumber
\\
\phi^2(R+kV+kL_M)]\nonumber
\eea
We are interested in studying flat Minkowski background space so $g_{LG}=\eta_{LG}$ and $|g|=1$ and $R=0$. So the geometric action
is again
\bea{}
S\left(\phi\right)=\int dt\int dx^{3n}\left(\frac{12n}{1-3n}(\partial^L\phi)(\partial_L\phi)+\phi^2(kV+kL_M)\right)\nonumber,
\eea
where the appearance of the potential $V$ was imposed externally.

From this action, the equation of motion for $\phi$ is
\bea{}
\frac{12n}{1-3n}\frac{\partial^L\partial_L\phi}{\phi}=k(V+L_M)\nonumber
\eea
with
\bea{}
L_M=\frac{2(\partial^L\tilde{S}_H)(\partial_L\tilde{S}_H)}{2\hat{M}_G}+\frac{\partial\tilde{S}_H}{\partial t_1}-\hat{M}\nonumber
\eea
and thus
\bea{}
\frac{12n}{k(1-3n)}\frac{\partial^L\partial_L\phi}{\phi}\!=\!\frac{2(\partial^L\tilde{S}_H)(\partial_L\tilde{S}_H)}{2\hat{M}_G}\!+\!\dot{\tilde{S}}_H \!+\!V\!\!-\!\hat{M}.
\label{twenty eight}
\eea
With the defined matching conditions, \eqref{twenty eight} is identical to \eqref{twenty seven}.

\section{\textbf{Discussions}}

In context with the above relation one should mention a few points
\begin{itemize}
\item Locality:\\
Any geometric description in analogy to the language of GR is supposed to be local,
where the rewritten quantum equations (\ref{nine}-\ref{twelve}) are clearly non-local.
``How, could such descriptions be equivalent?''\\
The answer to this question comes from the fact that the coordinates $x^L$ of the (local) geometric description
are defined in the higher dimensional configuration space of the system, while
the coordinates $\vec x$ in the (non-local) quantum description are initially defined in three dimensional space
plus a universal time.
\item Is the presented geometric description ``gravity''?\\
Even thought, the geometric description uses a covariant metric formalism,
it is not gravity, it is just a geometrical rewriting of the same quantum mechanical theory.
The distinction from gravity comes
due to several reasons. \\
First, it is defined in configuration space and not in 4-D spacetime.
Second, only the covariant derivatives of the
conformal factor $\phi$ on a flat background is contemplated and off-diagonal contributions are neglected,
like it would be done in a perturbation theory ansatz.
Third, the matter Lagrangian and its coupling constant differ from the usual
coupling between gravity and matter.
\item Matching conditions:\\
A set of matching conditions is defined to connect non-relativistic quantum mechanics with the classical geometric theory. These matching conditions are not unique but are chosen depending upon the situation, so as to provide a suitable connecting link between the two theories. These conditions connect the quantum phase $\tilde{S}$ with the Hamilton principle function $\tilde{S}_H$, the amplitude of pilot wave $P$ with the conformal function of the metric $\phi$ and the mass $M_j$ with the mass $\hat{M}_G$.
The coupling constant in the geometrical reformulation is given by
\bea{}
k = \frac{12n}{1-3n}.\frac{2\hat{M}_G}{\hbar^2}\nonumber,
\eea
which depends on the number of particles e.g. for
$n=1$, $k = \frac{-12\hat{M}_G}{\hbar^2}$.
\end{itemize}

\section{\textbf{Summary}}

It has been shown that,
for spinless particles, the $n$-particle equations of non-relativistic quantum mechanics can be
rewritten in the language of a geometric theory in the $1+3n$ dimensional configuration space.
 A set of matching conditions is defined to translate one formulation to the other.
 We further generalized the study for bosons subjected to a potential~$V$.

It is hoped that this work on the
many particles Schr\"odinger equation allows the interested reader
to add a different, a geometric, perspective on the laws of quantum mechanics.


\end{document}